\documentclass[a4paper,fleqn]{cas-dc}

\usepackage[numbers]{natbib}
\usepackage{lmodern}

\usepackage{caption}
\usepackage{xcolor}
\usepackage{algorithm}
\usepackage{algorithmic}
\usepackage{comment}

\usepackage{amsmath}
\usepackage[algo2e]{algorithm2e}

\usepackage{array}

\usepackage{enumitem}
\usepackage{etoolbox}
\usepackage{tikz}

\newcommand{\circled}[1]{
  \tikz[baseline=(char.base)]{
    \node[shape = circle, draw, inner sep = 1pt]
    (char) {\phantom{\ifblank{#1}{#1}{#1}}};
    \node at (char.center) {\makebox[0pt][c]{#1}};}}
\robustify{\circled}

\usepackage{enumitem}

\def\tsc#1{\csdef{#1}{\textsc{\lowercase{#1}}\xspace}}
\tsc{WGM}
\tsc{QE}
\tsc{EP}
\tsc{PMS}
\tsc{BEC}
\tsc{DE}
\begin{document}
\sloppy

\let\WriteBookmarks\relax
\def\floatpagepagefraction{1}
\def\textpagefraction{.001}
\shorttitle{Fast and Scalable Authentication Scheme}
\shortauthors{Li et~al.}

\title [mode = title]{A Fast and Scalable Authentication Scheme in IoT for Smart Living}

\author[add1]{Jianhua Li}[orcid=0000-0001-5878-6032]
\ead{jackleeml@hotmail.com}
\cormark[1]
\cortext[cor1]{Corresponding Author}

\author[add2]{Jiong Jin}
\ead{jiongjin@swin.edu.au}
\author[add3]{Lingjuan Lyu}
\ead{lyulj@comp.nus.edu.sg}
\cormark[1]
\author[add4]{Dong Yuan}
\ead{dong.yuan@sydney.edu.au}
\author[add5]{Yingying Yang}
\ead{yingying.yang@uts.edu.au}
\author[add1]{Longxiang Gao}
\ead{longxiang.gao@deakin.edu.au}
\author[add7]{Chao Shen}
\ead{chaoshen@mail.xjtu.edu.cn}

\address[add1]{School of Info Technology, Deakin University, Melbourne, VIC 3125, Australia.}
\address[add2]{School of Software and Electrical Engineering, Swinburne University of Technology, Melbourne, VIC 3122, Australia.}
\address[add3]{The National University of Singapore, Singapore.}
\address[add4]{School of Electrical and Information Engineering, The University of Sydney, Sydney, NSW 2006, Australia.}
\address[add5]{School of Electrical and Data Engineering, The University of Technology Sydney, Sydney, NSW 2007, Australia.}
\address[add7]{School of Electronic and Information Engineering, Xi'an Jiaotong University  Xi'an, SHX 710049 China.}

\begin{abstract}
Numerous resource-limited smart objects (SOs) such as sensors and actuators have been widely deployed in smart environments, opening new attack surfaces to intruders. The severe security flaw discourages the adoption of the Internet of things in smart living. In this paper, we leverage fog computing and microservice to push certificate authority (CA) functions to the proximity of data sources. Through which, we can minimize attack surfaces and authentication latency, and result in a fast and scalable scheme in authenticating a large volume of resource-limited devices. Then, we design lightweight protocols to implement the scheme, where both a high level of security and low computation workloads on SO (no bilinear pairing requirement on the client-side) is accomplished. Evaluations demonstrate the efficiency and effectiveness of our scheme in handling authentication and registration for a large number of nodes, meanwhile protecting them against various threats to smart living. Finally, we showcase the success of computing intelligence movement towards data sources in handling complicated services.
\end{abstract}

\begin{graphicalabstract}
\includegraphics{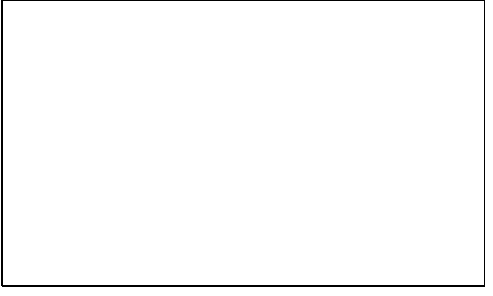}
\end{graphicalabstract}

\begin{highlights}
\item Investigating certificate authority as a service (CAaaS) concept grounded in microservices and fog computing.
\item Designing an effective and efficient CAaaS scheme for the authentication of the Internet of things (IoT).
\item Validating our proposed scheme via lightweight protocols.
\item Facilitating the mutual authentication for IoT endpoints to premise end-to-end security.  
\item Showcasing the success of computation movement toward data sources in handling complicated services.
\end{highlights}


\maketitle

\section{Introduction}

The Internet of things (IoT) has changed the world in many aspects of our living. For example, a user may install a Google smart speaker for less than \$100, performing voice recognition, information retrieval, and environment control \cite{Google2017speaker}. IoT-enabled applications such as healthcare and energy have already stepped into smart living \cite{Li2015ehopes}, bringing a variety of benefits including automation, efficiency, and low cost. Unfortunately, such applications are forming isolated information silos without enough cross-application communication and inter-cloud collaboration. With significant advantages of low latency, situational awareness, low cost, mobility support, and widespread coverage \cite{RAPUZZI2018building}, fog computing (fog) facilitates communication and collaboration among pervasive and ubiquitous things. As shown in Figure 1, fog is deployed in multiple tiers along the cloud-fog-things continuum, delivering substantial intelligence to where it is needed. However, widespread cyberthreats discourage people from further adoption of IoT devices in smart living \cite{KLOBAS2019how}.

\begin{figure}
\center
\advance\leftskip-0.5cm
    \includegraphics[width=0.8\columnwidth]{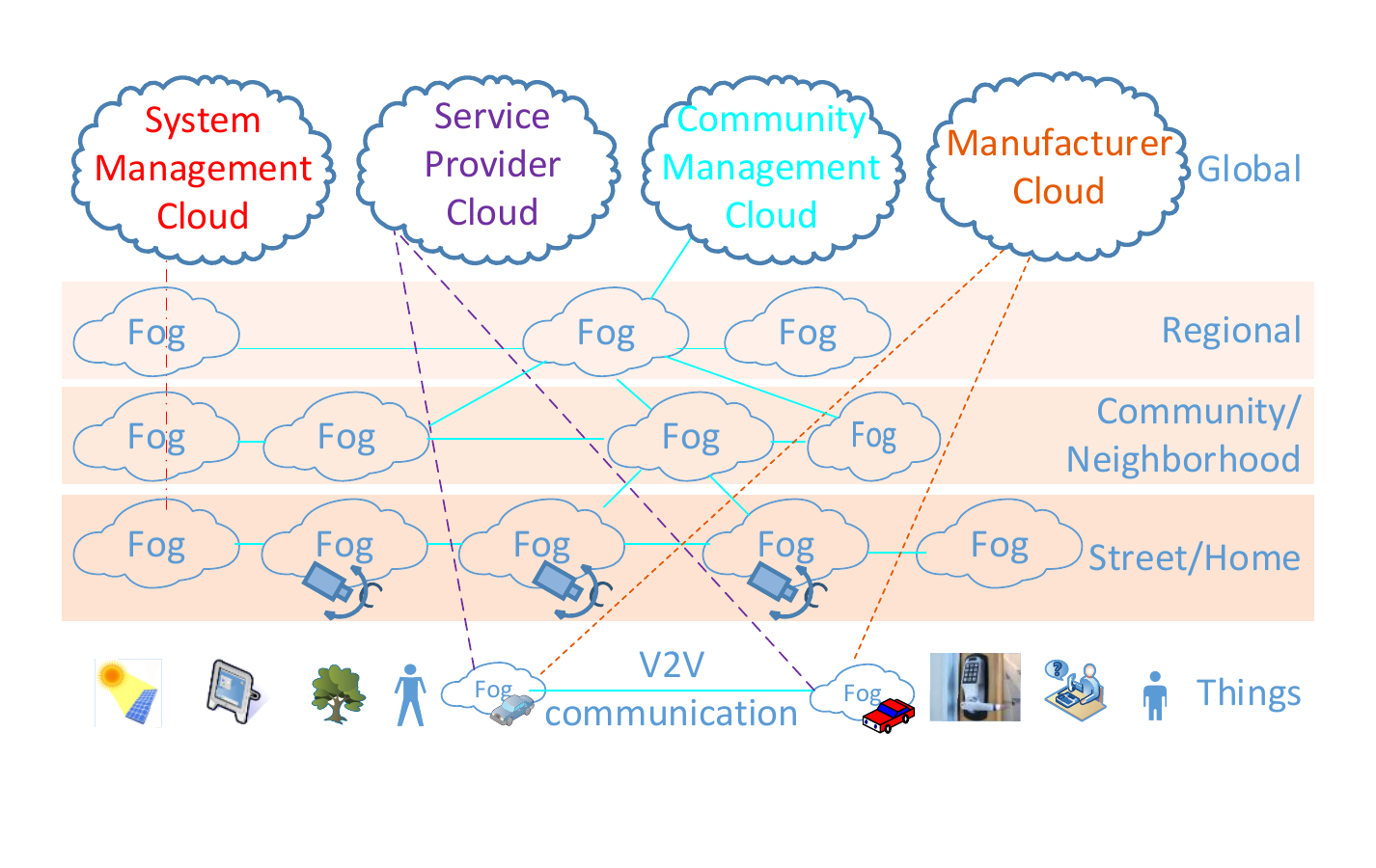}
    \caption{The cloud-fog-things continuum for smart living}
    \label{f1}
\end{figure}

The current combination of cloud and IoT is a game-changer technology that addresses many problems, although it is associated with considerable security deficiencies. Without mature legislation in the IoT market, profit-driven vendors often prioritize functionality over security in designing a product to save time and money, opening numerous attack surfaces. According to the U.S. Department of Justice, there are more than 4000 ransomware attacks each day on average since 2016. With little or no effort, an attacker can hack into home cameras to watch anything within the scope and capture videos and pictures \cite{liranzo2017security}. The deployment of such security-flaw devices will cause catastrophic disasters in many aspects.

 It is compulsory to recognize all vulnerabilities and then implement reliable solutions to minimize potential risks and achieve end-to-end security \cite{Neshenko2019survey}. Such solutions may involve computation-intensive workloads that are not feasible on many resource-limited SOs. Indeed, they have the same security requirements as powerful devices because any compromised device may potentially undermine the entire security pillar. Therefore, numerous efforts have been made in addressing the above challenges by using various technologies \cite{Salami2016Lightweight}, \cite{Almadhoun2018blockchain}, \cite{yan2016security}, \cite{AHMED2017bigdata}, \cite{WAZID2020lam}, \cite{KHATTAK2019perception}, \cite{shu2016security}.

\begin{table}[ht!]
\label{T1}
\caption{Abbreviation and symbol used in this paper}
\resizebox{0.9\columnwidth}{!}{
\begin{tabular}{|l|l|}
\hline
\multicolumn{1}{|c|}{\begin{tabular}[c]{@{}c@{}}Abbreviation\\ / Symbol\end{tabular}} & \multicolumn{1}{c|}{Explanation}                      \\ \hline
AI                                                                                    & Artificial intelligence                                                                                                                                                  \\ \hline
CA                                                                                    & Certificate authority                                                                         \\ \hline
CAaaS                                                                                    & Certificate authority as a service                                                                       \\ \hline
CRL                                                                                   & Certificate revocation list                                                                                    \\ \hline
ECC                                                                                   & Elliptic curve cryptosystem                                                                                    \\ \hline
FN                                                                                    & Fog node                                                                                                     \\ \hline

MFA                                                                                    & Multi-factor authentication                                                                               \\ \hline
ML                                                                                    & Machine learning                                                                                           \\ \hline
MSB                                                                                   & Microservice block                                                                                             \\ \hline
NFVM                                                                                  & Network function virtualization manager                                                                       \\ \hline
PKI                                                                                   & Public key infrastructure                                                                                      \\ \hline
SDN                                                                                   & Software-defined networking                                                                                    \\ \hline
SO                                                                                    & Smart object                                                                                                   \\ \hline
SVM                                                                                   & Service virtualization manager                                                                                 \\ \hline
VNF                                                                                   & Virtual network functions                                                                                      \\ \hline
VO                                                                                    & Virtual object                                                                                                 \\ \hline
$\times$                                                                              & Multiply, encryption, decryption                                                                               \\ \hline
$A_{ID_C}$                                                                            & An ID-based authentication key for client C                                                                    \\ \hline
$C, C_i, C_j$                                                                         & A client or child                                                                                              \\ \hline
$ID_C$                                                                                & An ID of client C                                                                                              \\ \hline
$E_i, E_j, E_{k_ij}, E_N$                                                             & An encrypted message                                                                                           \\ \hline
$IVV, IVV*$                                                                           & An integrity verification value                                                                                \\ \hline
$k, k_{ij}, k_{iS}, k_{jS}$                                                           & A symmetric session k                                                                                         \\ \hline
$K_S^R$                                                                               & A private key of CA server                                                                                    \\ \hline
$K_S^U$                                                                               & A public key of CA server                                                                                     \\ \hline
$M_C$                                                                                 & A temporary message based on ECC for a client                                                                 \\ \hline
$M_k, M_k'$                                                                           & A temporary message based on a session key                                                                    \\ \hline
$M_S$                                                                                 & A temporary message based on ECC for a client                                                                 \\ \hline
$N, N_i, N_j, N_j'$                                                                   & A nonce                                                                                                        \\ \hline
$P, Q$                                                                                & An initialized starting point on an ECC curve                                                                  \\ \hline
$Q_{ID_C}$                                                                            & \begin{tabular}[c]{@{}l@{}}Calculated coordinates of point Q from P on\\ an ECC curve\end{tabular}           \\ \hline
$R_C, R_C^*, R_C', R_C''$                                                             & \begin{tabular}[c]{@{}l@{}}The coordinates of client C on an ECC curve\\ with a different timestamp\end{tabular} \\ \hline
$R_S, R_S'$                                                                           & \begin{tabular}[c]{@{}l@{}}The coordinates of server S on an ECC curve\\ with a different timestamp\end{tabular} \\ \hline
$S$                                                                                   & A server or a device offers CA services                                                                        \\ \hline
$T_1,T_1', t_1, T_2$                                                                  & \begin{tabular}[c]{@{}l@{}}A different timestamp for that moment when an\\ event starts\end{tabular}          \\ \hline
$x_C, x_C'$                                                                           & The x coordinate of client C on the ECC curve                                                                  \\ \hline
$x_S$                                                                                 & The x coordinate of server S on the ECC curve                                                                  \\ \hline
\end{tabular}}
\end{table}

It is worth noting that security starts with reliable authentication to ensure the identification legitimacy of participating parties, which is a prerequisite for other security services like authorization and access control \cite{jing2014security}. In smart living, users (via an endpoint) and things must be mutually authenticated for secure communication, preventing it from executing a malicious process, data leakage, and privacy breach \cite{lyu2018ppfa}. As reviewed in \cite{sheng2013survey}, \cite{KHAN2018blockchain}, people have developed a diversity of standards and protocols in this direction. 

Due to a lack of a uniformed industrial standard for security among heterogeneous IoT nodes, the public-key-infrastructure (PKI) tends to be employed to provide authentication services \cite{Anderson2014looking}, where an entity owns a pair of public and private keys. In this system, a public key is presented as a form of a digital certificate to integrate its identity, while a private key manifests the ownership of the public key. Hence, the key pair of each entity must be effectively generated, managed, stored, and accessed through PKI. In this regard, a certificate authority (CA) is a centralized and trusted component to issue, store, and revoke keys if required.

Admittedly, the CA-based PKI works on many conditions, including robust endpoints, stable connections, and centralized management, but heterogeneous and ubiquitous IoT endpoints \cite{Kafhali2017efficient} may invalidate such assumptions in many ways. First, asymmetric cryptography is impractical to use with numerous sensors due to resource limitations. Furthermore, centralized CA needs to maintain a big database of keys material for each entity, which causes long delays in the keys acquisition, authentication failure, and even more severe problems \cite{HUANG202010effective}. Even worse, the long-distance and big-volume data movements from ubiquitous things to the remote cloud provide excessive attack surfaces for crackers \cite{lyu2020cloud}, deteriorating the security situation in all aspects \cite{zhou2017security}. For these reasons, we consider moving the security computation closer to IoT endpoints, thus making improvements to address such issues.

Therefore, we are motivated to design a novel scheme called ``CA as a Service'' (CAaaS) by following the microservice architecture \cite{Fadda2019microservices} and the virtual fog framework proposed in \cite{Li2018vfog}. In this scheme, we customize CA functionality for the IoT and push it closer to IoT to cope with the above concerns. The contributions are fivefold.
\begin{itemize}
    \item Initially, we design a microservice-enabled and fog-based CA scheme addressing both strong security and low latency.
    \item Secondly, we develop lightweight protocols to validate our scheme, where there is no need to run complicated asymmetric cryptography algorithm on data-source devices. 
    \item Thirdly, we facilitate the realizing of short-lived key pairs to reduce the burden of CA servers and enhance the device-to-device security. 
    \item Meanwhile, we propose a situation-awareness approach to identify compromise in IoT nodes and then explore cognitive measures to minimize attack surfaces.
    \item Eventually, we showcase the success of the computation movement from centralized servers to network edges.
    \end{itemize}
    
The road-map of this paper is as follows. In the next section, we present an overview of our proposed scheme. We elaborate on its implementation in Section 3, followed by the experimental evaluation in Section 4, which showcases low latency and high scalability. In Section 5, we discuss the protection intelligence against diverse attacks, accompanied by a formal verification with AVISPA to prove the effectiveness. In Section 6, we present the related work. Furthermore, Section 7 pinpoints limitations and future directions, while Section 8 concludes the paper. Table 1 presents the abbreviations and symbols used in this paper for better readability.

\section{The Mutual Authentication Scheme Overview}

In the first place, we introduce the virtual fog (VFOG) for IoT. Next, we inaugurate the scheme in the framework.

\subsection{The VFOG Framework for IoT}

Fueled by virtualization technologies encompassing service virtualization, network function virtualization \cite{sun2019latency}, \cite{sun2019energy}, and object virtualization, the VFOG enables the fog's realization as investigated in \cite{Li2018vfog}, underpinning the majority of beneficial features. This framework includes three layers, i.e., a service virtualization layer, a network function virtualization layer, and an object virtualization layer from top to bottom, along the cloud-to-things continuum.

In the service virtualization layer of the framework, through managing and coordinating \textit{microservice blocks (MSB)}, the service virtualization manager (SVM) empowers service orchestration, service monitoring, and service optimization based on MSBs, running on a pervasive FN. The MSB is a micro-service fragmenting application previously found on large servers \cite{thones2015microservice}. Section 2.4.2 reviews more details regarding MSB components.

In the middle, the network function virtualization layer decouples software network functions from vendor-specific equipment in the style of virtual network functions (VNFs). Such VNFs are immediately instantiated on an FN to facilitate networking functions among entities. Coupled with software-defined networking (SDN) technologies, the network function virtualization manager (NFVM) provisions, manages, coordinates, monitors VNFs, and its underlying infrastructure, enforcing resiliency and security.

At the bottom object virtualization layer, a virtual object (VO), hosted on an FN, seamlessly integrates an IP-enabled and sensor-specific network domain. On the one hand, it enhances the capability of an SO by taking over some resource-intensive workloads such as security. On the other hand, it manages and controls SOs, therefore further augmenting security with cognitive working intentions.

In the overarching VFOG framework, a lightweight middleware called foglets is responsible for the interplay between FN and clients. Apart from this, the VFOG employs foglets to manage the network infrastructure and empower user interface at endpoints, grounded with local situational awareness.

Most interestingly, the VFOG adds more benefits in the smart living transition, that is, transforming traditional devices and existing applications smoothly into a future environment. In consequence, the adoption of this framework makes smart living more finance-viable and affordable. To put it differently, the VFOG could protect investments made previously in an evolutionary process, although it is indeed a revolutionary reconstruction.

Despite the above benefits, this framework brings numerous advantages like low latency and cost. Security is also enhanced grounded on at least two facts initiated by VO, namely, heavyweight security workloads offloaded from an SO, and flexible management imposed on an SO with local awareness of working context. To this end, our scheme concentrates on security improvement in alignment with the two facts.

It is also worth noting the VFOG enables the realization of cloud-to-things continuum systematically with increased situational awareness along the continuum. Before reviewing our proposed authentication scheme, we first present the problem description and threat model as follows.

\subsection{Problem Description}
Although the VFOG framework is a key to the fulfillment of the cloud-fog-things continuum \cite{Li2018vfog}, the IoT security is still a challenge before practical adoption. Therefore, our primary goal is to explore an effective (i.e., against various attacks) and efficient (i.e., low-latency and scalable) authentication scheme, particularly suitable for resource-constrained devices.

Manufacturer clouds store detailed data of their products, including serial numbers, firmware, ID, software, embedded security feature, and other detailed specifications. Vendors conduct their business based on such critical data, supplying devices and applications to the market. At this moment, users order devices and services provided along the cloud-fog-things continuum. To generalize our description, we define fog as the \textit{Custodian} that has the responsibility to protect things (sensors, actuators, endpoints, etc.). Comparatively, a manufacturer cloud is the  \textit{Parent}, while things are the \textit{Child}. Again, manufacturers have detailed knowledge of their products, which is the \textit{Parent-Child affinity}. Empowered by such affinities, a \textit{Custodian} accomplishes situational protection for the IoT with local security awareness and working intention awareness.

As illustrated in Figure 2, a \textit{Child} is adopted in a cloud-fog-things continuum. From the perspective of security, it is feasible to achieve data security between \textit{Parent} and \textit{Custodian} (green flow) by collaboratively using advanced technologies. It is, nevertheless, inappropriate to a \textit{Child} due to various constraints, including energy, memory, computational speed, and network capability. Therefore, we focus on the authentication service provisioning for things (red flow). To this end, there are two questions raised, i.e., what scheme is suitable for the resource-constrained environment and how to implement it. In this paper, we first explore mutual authentication and symmetric key agreement between \textit{Child} and \textit{Custodian}. Then, we study inter-\textit{Child} authentication and key management via a trusted \textit{Custodian}.

\subsection{Threat Model}

As a general rule, a device in the cloud is securely protected rather than a fog node on-premises \cite{Cloud2019secure}. Due to actual circumstances (computing power, hardware, and software feature, deployment surroundings, business purpose, etc.), some FNs can take benefit from cutting-edge technological combinations for security. As such, it is acceptable to assume that some FNs are securely protected, while other FNs and IoT nodes are not in the continuum. In smart living, insecure nodes are susceptible targets to many cyber-physical attacks, including physical attacks (modifying a node with untrusted hardware/software components), passive attacks (eavesdropping and traffic analysis), replay attacks (pretending to be a legitimate user by capturing some valid data), and impersonation attacks (assuming the legal identity of an alternative communication participant).

In theory, CA distributes key services from a centralized server to CA proxies in a \textit{Custodian} device along with the cloud-fog-things continuum. A secure FN can be a \textit{Custodian} node, which forms a secure black box with cloud servers, referring to the secure zone triangle (the green triangle) in Figure 2. A \textit{Child} needs to register to a \textit{Custodian}. During this process, a \textit{Custodian} performs an integrity check, authentication, and session key negotiation for a \textit{Child}. After this, it serves as a trusted point for inter-\textit{Child} authentication. We assume that full-fledged security technologies fairly guarantee data security on the \textit{Custodian} device.
 
Conversely, a \textit{Child} such as a non-\textit{Custodian} FNs or an IoT endpoint locates in an insecure zone (the red triangle in Figure 2), which is hackable in terms of software, firmware, hardware, and communication. An intruder can eavesdrop data between \textit{Child} and \textit{Custodian}, manipulate data as wanted (e.g., delete, inject, delay, duplicate, replay). It even can acquire legitimate session keys with a \textit{Custodian} to launch further attacks against data-in-transit among other devices in this zone.

\subsection{CA as a Service (CAaaS)}
Technically, the most critical role of a CA is to serve as a trust anchor in the IoT ecosystem through which certificates are issued, distributed, and revoked. In this section, we briefly explain the cloud-fog-things continuum to contextualize our scheme.

\subsubsection{Cloud-Fog-Things Continuum}
According to the OpenFog Consortium, fog distributes resources and services of computing, storage, control, and networking anywhere along the cloud-fog-things continuum \cite{Li2018vfog}. As shown in Figure 1, fog computing provides computing intelligence at a hierarchical location of regional, community/neighborhood, street, and home, with the cognition of working conditions. It is thus equitable to deploy such resources and services to where they are most in need. From the data's point of view, data are collected, filtered, aggregated, stored, and processed with appropriate resources and services within proximity of data source, which significantly reduces communication cost \cite{CHALLA2020design} and alleviates bandwidth bottleneck.

\begin{figure}
\center
\advance\leftskip-0.5cm
    \includegraphics[width=0.8\columnwidth]{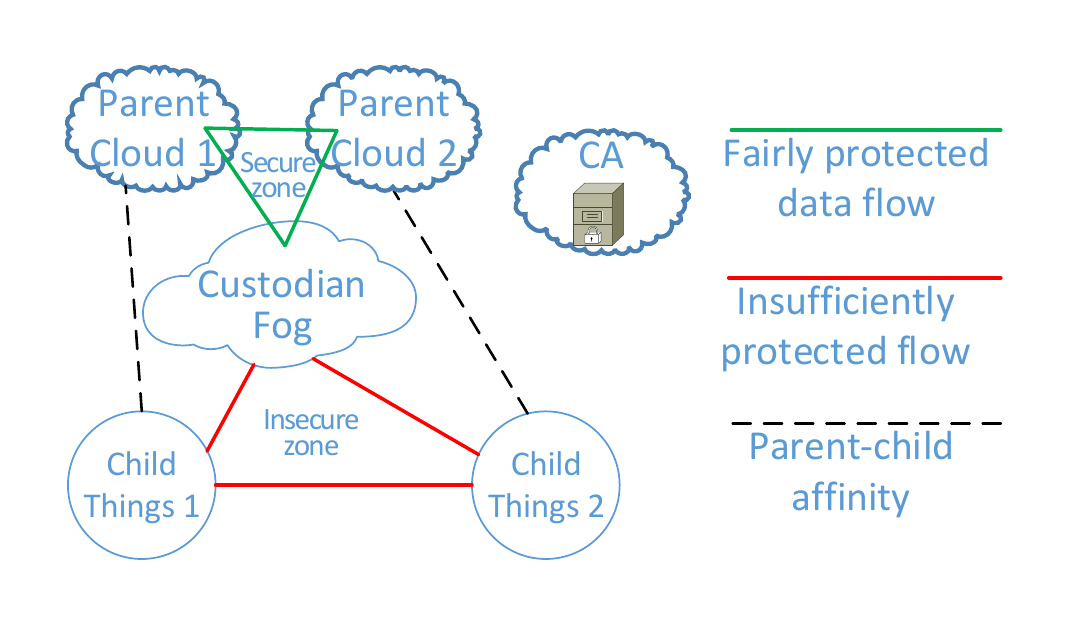}
    \caption{The secure and insecure area: There are two concerns of authentication raised for the insecure zone, namely, what scheme performs well and how to implement it?}
    \label{f2}
\end{figure}

The VFOG framework makes use of the technological landscape of cloud computing to make up the cloud-fog-things continuum. Relevant technologies incorporate resource pooling, resource management, service provisioning, multitenancy, load balancing, distributed storage, middleware, API, web service, and virtualization. Recent technological advancements in SDN, big data \cite{kaur2020energy}, deep learning \cite{garg2019depplearning}, artificial intelligence (AI), docker, microservice are also instrumental in bringing benefits such as low latency, high efficiency, scalability, and security \cite{lyu2019fog}, \cite{lyu2020foreseen}, \cite{zheng2019energy}.

As we focus on authentication in this paper, a detailed survey on such a technological integration is beyond the scope. Next, we explore MSBs to scrutinize CA components in the scheme.

\subsubsection{Microservice Block (MSB)}
As the trust anchor, a CA plays critical roles in policy authority, certificate issuer, certificate generator, revocation manager, registration authority, repository, and authentication officer. A CA runs on a powerful server, traditionally, sitting in a well-protected cloud data center. However, it is hard for mobile devices to detect such a CA server in a dynamic IoT ecosystem. For example, mobile nodes like driving cars would likely incur quick network topology change, while a blind effort to increase the numbers of Child CAs easily enkindles a security breach.

As demonstrated in \cite{tonella2003using}, it is achievable to break down large programs into small slices \cite{newman2016slice}, based on which we can perform the same function, in distributed nodes. We, therefore, take the advantage to decompose previous large programs into current MSBs. An MSB is a segmented software instance bearing discrete, autonomous, and network-accessible services. Each MSB is loosely-coupled and highly cohesive, which independently deploys on a widespread fog hardware platform. Such an MSB is a single application in light of working intentions.

A CA server in a cloud is well-protected by cutting edge security implementations in all aspects (logically, electronically, physically, etc.). In our scheme, we employ MSBs to bear the functions of a traditional CA, including generation, storage, and distribution of keys, certificate revocation list (CRL) services, CA proxy services, and more. Most importantly, evaluation results demonstrate that our scheme strengthens the security of CA. Next, we scrutinize the proposed CAaaS scheme composed of MSBs along with the cloud-fog-things continuum.

\subsubsection{CAaaS Scheme}

In brief, our CAaaS scheme has outstanding performance, resulting from the deployment of corresponding MSBs with situational awareness. A \textit{service virtualization manager (SVM)} manages such MSBs dynamically according to the actual authentication needs \cite{Li2018vfog}. To clarify, an SVM initializes, monitors, coordinates, and controls MSBs in the continuum to facilitate the provisioning of authentication services. Note that an MSB may or may not necessarily exist permanently at a particular \textit{Custodian} FN. Consequently, the local profile-based scheme provides unique benefits in authentication service provisioning.

\begin{figure}
\centering
\label{F3}
\advance\leftskip-0.5cm
\includegraphics[width=0.8\columnwidth]{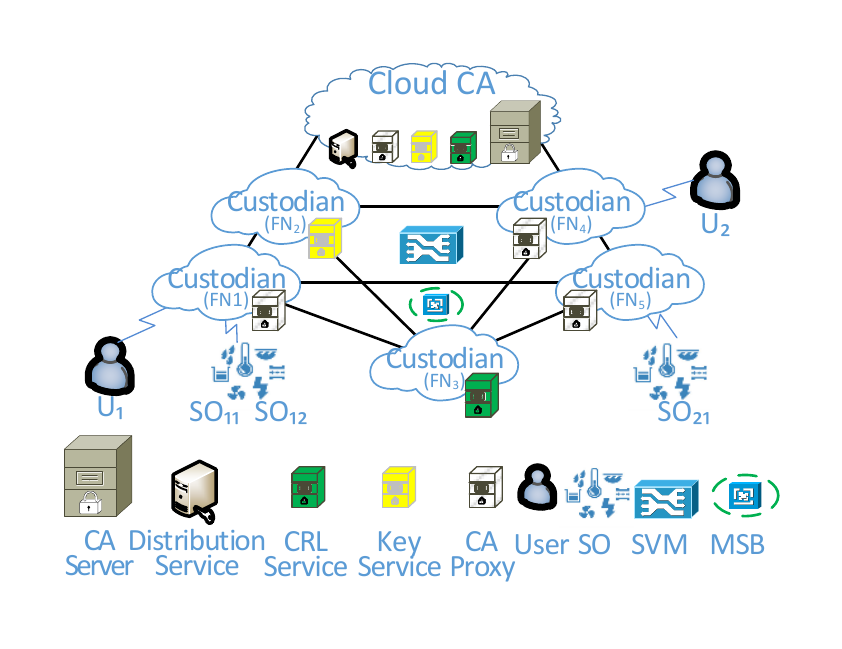}
\caption{An overview of our proposed CAaaS scheme.}
\end{figure}

As shown in Figure 3, the VFOG distributes CA MSBs to capable FNs along the cloud-fog-things continuum. Such an FN is a \textit{Custodian}, bearing CA services and protecting nearby IoT components.  Under the coordination of an SVM, these MSBs make up needed functions at the proximity of data source in fog \cite{Chiti2019reassemble}, which improves the availability, efficiency, and scalability of CA \cite{TARNEBERG2017placement}. To demonstrate, FN$_1$ performs the authentication between SO$_{11}$, SO$_{12}$, and user U$_1$.  

Provided that two entities are accessing the fog via different locations, for example, user U$_1$ and U$_2$, the positioning of authentications becomes non-trivial. Based on the latency estimation service provided by fog \cite{Li2017file}, a node with the least overall latency, such as FN$_2$, may fulfill the authentication between the two. From the perspective of FN's deployment, we dispatch keys services as well as CRL services on upper layers, keys distribution services in middle layers, and CA proxy services at lower tiers. In effect, the reduced distance diminishes latency and risk of data movements, fostering security, and scalability in the IoT ecosystem.

To further validate our scheme, We develop four protocols in the implementation. In the system configuration phase (detailed in Protocol 1  in Section 3.3), the CA generates its private and public key pair and publishes its public key to all networked nodes, getting ready for \textit{Child} registration. During the registration and authentication process (performed with Protocol 2), CA generates an ID-based authentication key and passes to the \textit{Child} in a secure channel. Simultaneously, a \textit{Custodian} completes device integrity verification (achieved in Section 3.2), taking advantage of the \textit{Parent-Child affinity}. Then, the \textit{Child} can use the authentication key for mutual authentication and key agreement with a \textit{Custodian} (accomplished with Protocol 3). After \textit{Custodian-Child} mutual-trust is established, the \textit{Custodian} is further used as a trust point for inter-\textit{Child} authentication and key management (fulfilled in Protocol 4).

It is also worth mentioning that some \textit{Custodian} may be able to issue short-lived keys for fast-moving things, e.g., driving cars. Location-based and situation-awareness authentication may be required to get local service, while the temporary key expires when the vehicle leaves the area. In this scenario, the \textit{CRL} service would be kept busy without the support of a short-lived key. In more detail, a CRL is an array of certificates that have been marked as "no longer trusted" before their scheduled expiration.

\section{The Scheme Implementation}
In this section, we review the preliminary of ECC cryptography, device integrity check, and proposed protocols.
\subsection{The Preliminary}

An elliptic curve $E_p(a,b)$ over a Galois field $F_p$, where $p>3$ and is prime, is the set of all ($x,y$) ($x,y \in F_p$) that satisfies the cubic equation:\\

$y^2=x^3+ax+b$  \hfill (1) \\

\noindent where $a,b \in F_p$, and $4a^3+27b^2 \neq0$. \\
The points on the curve with an extra point $O$ called the point at infinity, form a group \\

$G=\{(x,y):x,y \in F_p, E_p(x,y)=0\}\cup \{O\}$  \hfill (2)\\

\noindent $G$ is a cyclic additive group defined as below:\\
Let $P,Q \in G$, a line $l$ containing $P$ and $Q$ (tangent line to $E$ if $P=Q$), meets $E$ at $R$. Let $l'$ be a line linking $R$ and $O$. Then, $P$ ``+'' $Q$ is the point where $l'$ intersects $E$ at $R$ and $O$, and $P$ ``+'' $Q$. Scalar multiplication over $E_p$ can be calculated as\\

$s \times P=P+P+\cdots+P$ (s times) \hfill (3)\\

\noindent where $s\in F_p*$ is an integer, and a point $P\in E_p(a,b)$.
In general, the security of ECC depends on the following difficulties:

\begin{enumerate}
\item 
Discrete logarithm problem (DLP): given $G$'s generator $P$ and $Q=x \times P$ to calculate x.
\item
Computational Difie-Hellman problem (CDHP): given $G$'s generator $P$, $a \times P$, and $b \times P$ to compute $ab \times P$. 

\item
Elliptic curve factorization problem (ECFP): given two points $P$ and $Q$=$s\times P$+$t\times P$ over $E_p(a,b)$ for $s,t\in F_p^*$, to locate two points $s\times P$ and $t\times P$ over $E_p(a,b)$.
\end{enumerate}

%%%%%%%%%%%%%table 2 - two-column
\begin{table*}[t!]
\centering
\caption{Possible attacks and countermeasures}
\label{T2}
\resizebox{0.95\linewidth}{!}{
\begin{tabular}{|l|l|l|l|} 
\hline
\multicolumn{1}{|c|}{Attack Surface}                                                                                                      & \multicolumn{1}{c|}{Possible Attacks}                                                                                          & \multicolumn{1}{c|}{Possible Consequence}                                                                                                                                                                                                                                 & \multicolumn{1}{c|}{Counter-measures}                                                                                                                                                   \\ 
\hline
\begin{tabular}[c]{@{}l@{}}Processors(CPU, \\GPU, etc.) \end{tabular}                                                                      & \begin{tabular}[c]{@{}l@{}}Swap processors {}with \\flawed {}ones \end{tabular}                    & Device damage, backdoors                                                                                                                                                                                                                                                  & \begin{tabular}[c]{@{}l@{}}Serial number check, device \\ quarantine, trustworthy downgrade \end{tabular}                                                                            \\ 
\hline
\begin{tabular}[c]{@{}l@{}}ROM, EPROM, \\ functionality\\-specific {}module\\(e.g., sensing, \\crypto module) \end{tabular} & \begin{tabular}[c]{@{}l@{}}Utilization of \\compromised firmware, \\default access \end{tabular}                               & \begin{tabular}[c]{@{}l@{}}Device malfunction, {}unwanted\\functions, {}device ID, security\\certificate attacks, undesired \\access through default settings, {}\\compromised {}cryptosystem, etc. \end{tabular} & \begin{tabular}[c]{@{}l@{}}Firmware content check, device\\quarantine, firmware recovery,\\ {}upgrade, patches, ID and \\certificate attestation \end{tabular}            \\ 
\hline
External memory                                                                                                                           & \begin{tabular}[c]{@{}l@{}}Malicious {}data, \\Malware, e.g., Trojan \end{tabular}                               & \begin{tabular}[c]{@{}l@{}}Comprehensive security breach,\\ {}e.g., device malfunction, \\network performance degradation \end{tabular}                                                                                                                   & \begin{tabular}[c]{@{}l@{}}Device hardening, malicious code and\\ behavior detection, whitelisting, device {}\\quarantine, {}sandboxing, etc. \end{tabular}  \\ 
\hline
\begin{tabular}[c]{@{}l@{}}Input/output\\interfaces \end{tabular}                                                                         & \begin{tabular}[c]{@{}l@{}}Untrusted {}data or \\device connection \end{tabular}                                 & \begin{tabular}[c]{@{}l@{}}Attacks gaining access to IoT \\device \end{tabular}                                                                                                                                                                                           & \begin{tabular}[c]{@{}l@{}}Unneeded interfaces shutdown, \\disconnection/connection monitoring \end{tabular}                                                                          \\ 
\hline
\begin{tabular}[c]{@{}l@{}}Transceivers/\\network adapters \end{tabular}                                                               & \begin{tabular}[c]{@{}l@{}}Fake node {}attack, {}\\routing attack, etc. \end{tabular}              & \begin{tabular}[c]{@{}l@{}}Fake node or data inserted to interfere {}\\normal business, {} energy {}wasting, {}\\undesired routing {}behavior, DoS attack \end{tabular}                             & \begin{tabular}[c]{@{}l@{}}Device whitelisting/blacklisting,\\host attestation, device quarantine \end{tabular}                                                                       \\ 
\hline
\begin{tabular}[c]{@{}l@{}}Embed expansion \\slot \end{tabular}                                                                           & \begin{tabular}[c]{@{}l@{}}Untrusted card \\installation \end{tabular}                                                         & Unwanted functionality                                                                                                                                                                                                                                                    & \begin{tabular}[c]{@{}l@{}}Redundant slot {}deactivation,\\card {}verification \end{tabular}                                                                \\ 
\hline
\begin{tabular}[c]{@{}l@{}}Operating \\system (OS) \end{tabular}                                                                          & \begin{tabular}[c]{@{}l@{}}OS/runtime compromising,\\ malicious coding, etc. \end{tabular}                               & \begin{tabular}[c]{@{}l@{}}OS crash or lockup, compromised \\ cryptosystem, runtime, etc. \end{tabular}                                                                                                                                                                  & \begin{tabular}[c]{@{}l@{}}Sandboxing, hardening, \\ lockdown, whitelisting, patches, reset \end{tabular}                                                                              \\ 
\hline
\begin{tabular}[c]{@{}l@{}}Software \\utilities \end{tabular}                                                                             & \begin{tabular}[c]{@{}l@{}}Bugs, backdoors\\exploitation \end{tabular}                                                         & \begin{tabular}[c]{@{}l@{}}Unauthorized device {}access, resource\\utilization, compromised cryptosystem,\\device ID and certificate modification \end{tabular}                                                                                  & \begin{tabular}[c]{@{}l@{}}Sandboxing, hardening, \\ lockdown, whitelisting,  blacklisting \end{tabular}                                                                              \\ 
\hline
Data                                                                                                                                      & \begin{tabular}[c]{@{}l@{}}Data-in-transit attack, data-\\in-rest attack, data collection\\from rejected devices \end{tabular} & \begin{tabular}[c]{@{}l@{}}Password, and other sensitive data loss,\\privacy leakage, secret loss, {}finance\\loss, reputation loss, etc. \end{tabular}                                                                                & \begin{tabular}[c]{@{}l@{}}Using protected connection (e.g, virtual\\private network), user track, cryptosystem, \\ethical {}isolation, pentest, etc. \end{tabular}      \\ 
\hline
User                                                                                                                                      & \begin{tabular}[c]{@{}l@{}}Social engineering,\\malicious actors \end{tabular}                                                 & \begin{tabular}[c]{@{}l@{}}Sensitive information loss, {}\\network misconfiguration,\\unauthorized access to device, etc. \end{tabular}                                                                                                                    & \begin{tabular}[c]{@{}l@{}}Enabling situational {} awareness using\\a combination of technologies such as \\big data, machine learning, pentest, etc. \end{tabular}       \\ 
\hline
\begin{tabular}[c]{@{}l@{}}Other physical\\attacks \end{tabular}                                                                          & \begin{tabular}[c]{@{}l@{}}EMI, RFI attacks,\\property vandalism, etc. \end{tabular}                                           & \begin{tabular}[c]{@{}l@{}}Losses of {} device, and\\connection, {}etc. \end{tabular}                                                                                                                                                         & Out of the scope                                                                                                                                                                        \\
\hline
\end{tabular}}
\end{table*}

\subsection{Device Integrity Verification}
Working in an insecure zone, a \textit{Child} offers numerous attack surfaces, including processors, function-specific modules, operating systems, and software utilities. Thanks to fog enabling the situational awareness of manufacturers, service providers, users, and working surroundings where an SO operates, the VO performs vulnerability assessment and mitigation, based on available contexts. To illustrate, hackers can frequently scan and exploit system vulnerabilities to install backdoors, sniffers, data collection software, and even remotely control a victim to launch malicious attacks. Fortunately, a \textit{Custodian} can determine if a \textit{Child} integrity is compromised, through \textit{Parent-Child affinity}, and take reasonable actions surveyed in Table 2, therefore minimizing threats. 

Following the above action, the VO calculates the \textit{integrity verification value (IVV)} to verify whether there is any modification on a \textit{Child}. The parameter contributing to this value may include ID, firmware, OS, running software, used slots, unused slots, blacklisting services, and more. Meanwhile, the \textit{Custodian}  computes it (\textit{IVV*}) based on the parameter obtained from the affinity. If the two values match, there is no modification; otherwise, the \textit{Custodian} may reset the device or conduct device quarantine and inform other nodes of related security risks. As such, the system maintains consistent local security awareness. 

In more detail, a \textit{Child} device (user devices, sensors, etc.) needs to register to fog before performing its business functions. During the registration process, the \textit{Custodian} conducts verification and device hardening. A device may have preloaded security features such as encryption algorithms and pre-shared keys, through which establishing an exclusively confidential communication channel to transmit ID-based authentication keys. This isolation prevents an ID-based authentication key from being leaked.

\begin{table}[t!]
\label{T3}
\centering
\caption{Key size comparison for the same level of security}
\resizebox{0.9\columnwidth}{!}{
\begin{tabular}{|c|c|c|}
\hline
\begin{tabular}[c]{@{}c@{}}Symmetric Encryption\\ (key size in bits)\end{tabular} & \begin{tabular}[c]{@{}c@{}}RSA \& Diffie-Hellman\\ (module size in bits)\end{tabular} & \begin{tabular}[c]{@{}c@{}}ECC Cryptography \\ (key size in bits)\end{tabular} \\ \hline
56                                                                                & 512                                                                                   & 112                                                                            \\ \hline
80                                                                                & 1024                                                                                  & 160                                                                            \\ \hline
112                                                                               & 2048                                                                                  & 224                                                                            \\ \hline
128                                                                               & 3072                                                                                  & 256                                                                            \\ \hline
192                                                                               & 7680                                                                                  & 384                                                                            \\ \hline
256                                                                               & 15360                                                                                 & 512                                                                            \\ \hline
\end{tabular}
}
\end{table}

It is equally important to point out that device update (namely, software patches, firmware update, components upgrade, etc.) is compatible with the device integrity verification. Fog actively collects such updates from various hardware manufacturers and software developers to maintain its currency in the IoT ecosystem. In case that a \textit{Child} device is hard to upgrade directly from its \textit{Parent} network due to diverse reasons (unstable connection, working surroundings, business requirement, etc.), fog may serve as an agent to perform such updates. Next, we investigate a series of protocols for the scheme implementation.

\subsection{The Protocol}
Grounded in the device integrity verification and device hardening, our proposed protocol achieves end-to-end authentication for IoT devices in the continuum. In brief, Protocol 1 initializes the system; Protocol 2 performs registration to CA; Protocol 3 conducts mutual authentication, generating a session key between user/device and CA, while Protocol 4 carries out the endpoint to endpoint authentication. To simplify the description, we name CA and its distributed service in the secure zone as server $S$ and an outside node as client $C$.

\begin{algorithm}[H]

\floatname{algorithm}{Protocol 1}
\renewcommand{\thealgorithm}{}
\caption{System settings and announcement}
\label{protocol1}
 \algsetup{linenosize=\small}
\scriptsize
\begin{algorithmic}[1]
\STATE $S$: Chooses an elliptic curve $E_p(a, b)$ over a large Galois field F and a base point $P$, on that curve as described in Section3.1. After this, it derives its private/public key pair ($K_S^R$, $K_S^U$), \\where $K_S^U$=$K_S^R \times P$.\\

\STATE $S$: Selects three secure hash functions: \\
$H_1(\cdot):\{0,1\}\rightarrow G_p$ \\
$H_2(\cdot):\{0,1\}\rightarrow Z_p^*$\\
$H_3(\cdot):\{0,1\}^*\rightarrow Z_p^*$ \\ 
where $G_p$ is generated by $P$ over $E_p(a,b)$ (referring to equation 2)

\STATE $S\rightarrow{C}$: Keeps $K_S^R$ in secret and broadcasts clients of\\
$\{E_p(a,b), P, K_S^U, H_1(\cdot), H_2(\cdot), H_3(\cdot)\}$

\STATE $S$: An SVM creates MSBs for keys distribution service, certificate revocation list (CRL) service, keys computation, storage service, and CA proxy service, and deploys these services on FN as required.
\end{algorithmic}
\end{algorithm} 

The announcement made by $S$ goes along with the cloud-fog-things continuum. Finally, $C$ notices the availability of CA. In this case, $C$ treats CA proxy as $S$, as it is kept back from nodes out of the secure black box. Then, a client starts its registration process through Protocol 2.

\begin{algorithm}[H]

\floatname{algorithm}{Protocol 2}
\renewcommand{\thealgorithm}{}
\caption{\textit{Child} registration to CA}
\label{protocol2}

 \algsetup{linenosize=\small}
\scriptsize
 
\begin{algorithmic}[1]
\STATE $C\rightarrow{S}$: Sends $\{ID_C\}$ via CA proxy\\
\STATE $S$: Computes the authentication key A$_{ID_C}$ for $C$: \\
A$_{ID_C}$=$K_S^R\times H_1(ID_C)\in G_p$
\STATE $S\rightarrow{C}$: Sends $\{A_{ID_C}\}$ via CA proxy in a confidential communication channel.\\
\STATE $C$: Computes A$_{ID_C} \times P$=$K_S^U\times H_1(ID_C)$. 
\\If true, keeps $A_{ID_C}$ in private. \\
\end{algorithmic}
\end{algorithm}

After registration, the \textit{Custodian} and \textit{Child} will authenticate each other and share a session key, using Protocol 3.

\begin{algorithm}[H]
\floatname{algorithm}{Protocol 3}
\renewcommand{\thealgorithm}{}
\caption{\textit{Custodian-Child} authentication}
\label{protocol3}

 \algsetup{linenosize=\small}
\scriptsize

\begin{algorithmic}[1]
\STATE $C$: Randomly selects a point $R_C=(x_C, y_C)\in E_p(a, b)$,\\ 
where $x_C$ and $y_C$ are coordinates of the point on x and y axis.\\
\STATE $C$: Generates a current timestamp $T_1$, and computes$t_1, M_C$, and $R_C^*$.\\
$t_1=H_2(T_1)$ \\
$M_C=R_C+t_1 \times A_{ID_C}$ \\
$R_C^*=X_C\times P$ \\
\STATE $C\rightarrow S$: Sends $\{ID_C, M_C, R_C^*, T_1\}$ via CA proxy.\\
\STATE $S$: Computes the coordinates of $Q_{ID_C}$ and $R_C'$. \\
$Q_{ID_C}=H_1(ID_C)=(x_Q, y_Q)$ \\
$t_1'=H_2(T_1)$ \\
$R_C'=M_C - t_1' \times K_S^R \times Q_{ID_C}=(x_C', y_C')$ \\
\STATE $S$: Generates the local current timestamp $T_2$, calculates $\Delta T$.\\
$\Delta T=T_2-T_1$\\
If $\Delta T$ is accepted, $S$ checks if $R_C^*=x_C' \times P$ holds. \\
If true, $S$ trusts the validity of $C$, and asserts $x_C'=x_C$.\\
Else, the protocol terminates.\\
\STATE $S$: Randomly chooses a point $R_S$: \\
$R_S=(x_S, y_S)\in E_p(a, b)$, and computes $t_2$ and $M_S$.\\
$t_2=H_2(T_2)$\\ 
$M_S=R_S+t_2 \times K_S^R \times Q_{ID_C}$\\
And then calculates a session key $k$ and temporary message $M_k$.\\
$k=H_3(x_Q, x_C, x_S)$\\
$M_k=(k+x_S)\times P$\\
\STATE $S\rightarrow{C}$: Sends $\{M_S, M_k, T_2\}$ to C via the proxy.\\
\STATE $C$: Computes the coordinates of $Q_{ID_C}$ and $R_S'$. \\
$Q_{ID_C}=H_1(ID_C)=(x_Q, y_Q)$\\ 
$t_2'=H_2(T_2)$\\
$R_S'=M_S-t_2' \times A_{ID_C}=(x_S', y_S')$\\
\STATE $C$: Verifies the delay is allowed by comparing $T_2$ with the current time, if true, $C$ then computes $k'$ and $ M_k'$\\ 
$k'=H_3(x_Q,x_C,x_S')$\\ 
$M_k'=(k'+x_S')\times P$\\ 
If $M_k'=M_k$, then $k=k'$ is true, which means the session key is successfully shared between $S$ and $C$.\\
Otherwise, the protocol is terminated. 
\end{algorithmic}
\end{algorithm}

Note that a client does not require bilinear-pairings, making the protocol more feasible on resource-constrained nodes. Grounded on the authentication and unique shared session keys, a \textit{Child} can now establish respective session keys with other nodes via any mutually trusted \textit{Custodian}, as demonstrated in Protocol 4.

\begin{algorithm}[H]
 
\floatname{algorithm}{Protocol 4}
\renewcommand{\thealgorithm}{}
\caption{Inter-\textit{Child} authentication and key management}
\label{protocol4}
 \algsetup{linenosize=\small}
\scriptsize

\begin{algorithmic}[1]
\STATE $C_i$: Proposes an encryption key $k_{ij} (i \neq j)$ and a nonce $N_i$ at random (will be used for authentication when the node is requested), initializing the key agreement, and calculates $E_j$,\\
$E_j = C_j \times k_{iS}$\\
where $C_j$ is the expected communication peer. Then, it encrypts the proposed key with its shared key$K_{iS}$.\\
$E_{k_{ij}} = k_{ij} \times k_{iS}$ \\
\STATE $C_i\rightarrow S$: Sends $\{E_j, E_{k_{ij}}\}$ to $S$.\\
\STATE $S$: Decrypts the received message from $C_i$ using the shared key $k_{iS}$, computes $C_j$ and $k_{ij}'$.\\
$C_j=E_j \times k_{iS}^{-1}$\\ 
$k_{ij}'=E_{k_{ij}'} \times k_{iS}^{-1}$ \\
\STATE $S$: Encrypts the request with shared key $ k_{jS}$.\\ 
$E_i = C_i \times k_{jS}$\\
$E_{k_{ij}'} = k_{ij}' \times k_{jS}$ \\
\STATE $S\rightarrow C_j$: Forwards $\{E_i, E_{k_{ij}}'\}$ to $C_j$.\\
\STATE $C_j$: Decrypts the received message from $S$ using the shared key $k_{jS}$.\\
$C_i = E_i \times k_{jS}^{-1}$,\\
$k_{ij}^{''} = E_{k_{ij}'} \times k_{jS}^{-1}$ \\
\STATE $C_j$: Generates a nonce $N_j$ and responds to $C_i$'s request. Uses the calculated session key $k_{ij}^{''}$ to encrypt the reply for authentication. \\
$E_j = C_j \times k_{ij}^{''}$\\ 
$E_{N_j} = N_j \times k_{ij}^{''}$\\ 
Meanwhile, $C_j$ memorizes $E_{N_j}$, waiting for authentication.\\
\STATE $C_j\rightarrow C_i$: Sends $\{E_j, E_{N_j}\}$ to $C_i$.\\
\STATE $C_i$: After receiving the reply, it uses its proposed key at step 1 to decrypt the message. Calculates $C_j$ and $N_j'$.\\
$C_j = E_j \times k_{ij} = E_j \times k_{ij}^{''}$\\
$N_j' = E_{N_j} \times k_{ij} = E_{N_j} \times k_{ij}^{''}$\\ 
When it holds, node $C_i$ can now talk to $C_j$ directly for further authentication.
\STATE $C_i$: Encrypts the nonce with the shared key$K_{ij}$, computes $E'_{N_j}$.\\
$E'_{N_j} = N_j' \times k_{ij}$ \\
\STATE $C_i\rightarrow C_j$: Sends $\{E_{N_j}'\}$ to $C_j$. \\
\STATE $C_j$: Compares the nonce received with stored.\\
If $E_{N_j}'$ = $E_{N_j}$, the authentication is performed.\\
Otherwise, the protocol is terminated.\\
\end{algorithmic}
\end{algorithm} 

In Protocol 4, $k_{iS}$ and $k_{jS}$ are symmetric session keys between client nodes $C_i$ and $S$, $C_j$ and $S$, respectively. Node $C_i$ and $C_j$ will be able to perform the negotiation of session keys via trusted $S$ and authenticate each other through a nonce $N$. When $C_i$ is going to communicate with $C_j$, node $C_i$ will initialize a session key $k_{ij}$, and send to $S$, where the key will be forwarded to $C_j$. After this, node $C_j$ generates a nonce $N_j$ and encrypts the nonce with key $k_{ij}$ to challenge $C_i$ for authentication.

As can be seen, Protocol 4 supplies the transitory session key negotiation and nonce-based authentication via a \textit{Custodian}. Based on this session key, \textit{Child}) devices can transact their session keys immediately for private communications. 

The above four protocols facilitate mutual authentication and key agreement between any two nodes through a mutually trusted \textit{Custodian}. In this process, the VFOG extends cloud-based CA authentication service to network edges. It is significant for delay-sensitive and fast-moving objects, for example, a moving car, to acquire traffic information along the street. In this case, our scheme issues a short-lived session key between the mobile client and a smart traffic system. 

Moreover, we take the \textit{Parent-Child affinity} into consideration for security. With the cognition of working intention and security context, $IVV$ contributes to the detection of modifications introduced by physical attacks, as later studied in Section 5. It is of significant importance to generate a symmetric session key for resource-limited devices because it has the smallest key size for the same level of security, as summarized in Table 3.

\section{The Scheme Evaluation}
As elaborated in Section 2, in the proposed scheme, we decompose computation-intensive CA-related services in a fashion of MSBs, then reassemble them to achieve the same function over multiple FNs. It is thus disseminating micro-services to the proximity of users and devices for authentication along the cloud-fog-device continuum. Since CA services are approaching IoT users and devices, leading to a reduced distance that decreases the registration and authentication delay, meanwhile improves the scalability. In this evaluation, we observe the latency depletion along with the cloud-fog-things continuum.

\begin{figure*}
\centering
%\advance\leftskip-0.5cm
\includegraphics[width=0.9\textwidth]{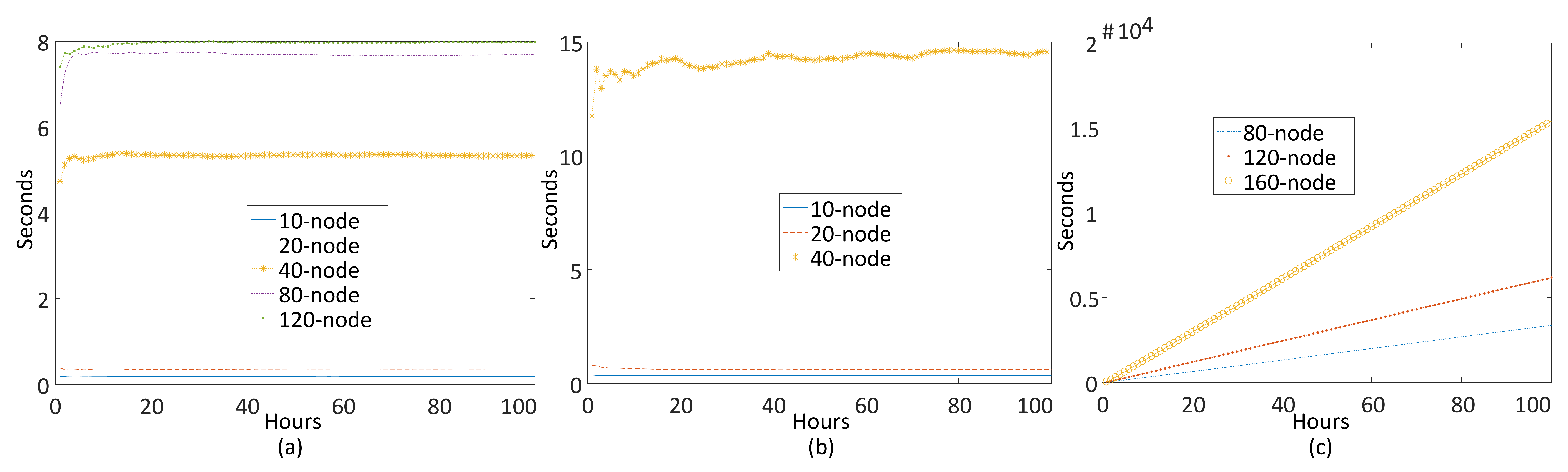}
\caption{Delay in the cloud-only model: (a) The average registration delay. (b) The authentication delay for a small number of nodes. (c) The authentication delay for a large number of nodes.}
\label{f4}
\end{figure*}

\begin{figure*}
\centering
%\advance\leftskip-0.5cm
\includegraphics[width=0.9\textwidth]{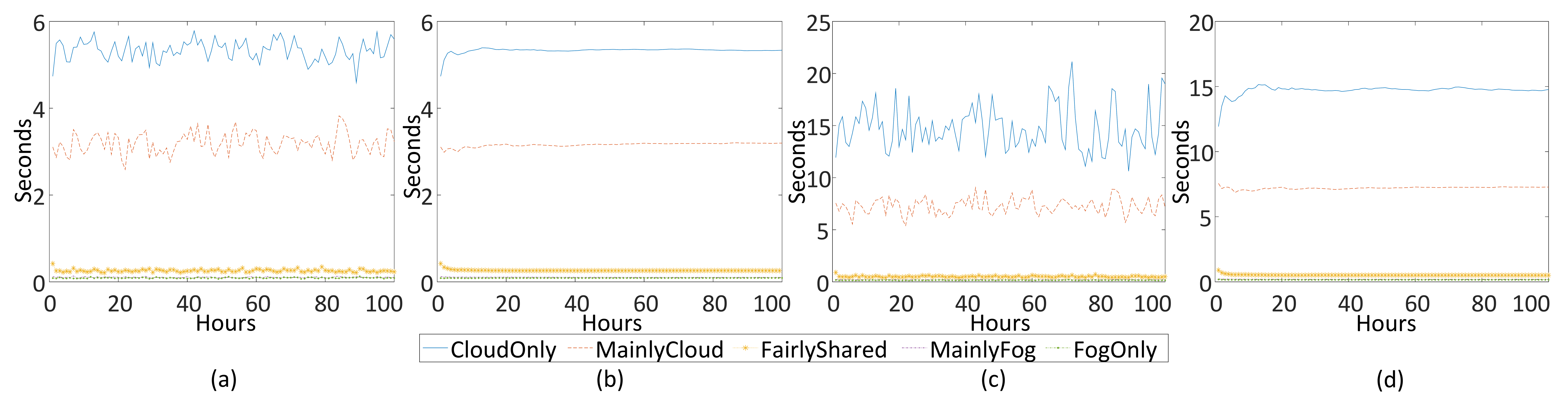}
\caption{Delay for a fair number of nodes in a shared model: (a) The instantaneous registration delay. (b) The average registration delay. (c) The instantaneous authentication delay. (d) The average authentication delay.}
\label{f5}
\end{figure*}

\subsection{Performance Index}
In this evaluation, we illustrate the advantage of our scheme, including low-latency of endpoint registration and authentication, as well as high scalability. First, we define the \textit{latency} as the end-to-end delay encompassing nodal processing delay and transmission delay measured in seconds. Second, we interpret the \textit{scalability} as the ability to expand in performing registration and authentication for growing numbers of transactions. The scalability is evaluated based on three parameters, namely, the average delay per transaction, workloads, and CPU utilization on servers. In more detail, the \textit{workload} refers to the number of tasks per second. Accordingly, the more pieces of work, the higher the workloads. 

Note that the evaluation of the two indexes are correlative, although their definitions are of fundamental differences. In general, for the same number of endpoints, the lower the average delay is, the higher the scalability it is. Similarly, when a system copes with a specific service, the less number of threads on CPU it is (i.e., fewer workloads), the more scalable it is (implying the potential to handle more tasks), on condition that the processor is not overwhelmed.

It is of the essence to harvest meaningful data in an evaluation process. We, therefore, select mainstream tools with proven effectiveness in network simulation \cite{SALAH2006opnet}. In this case, the OPNET Modeler and MATLAB are combined in use to collect the required latency and the CPU utilization data due. Particularly, OPNET Modeler is simulating both Cloud and Fog platforms, while the collected data are verified and validated by open-web tools \cite{leonhard2017cloudping} with the aid of MATLAB.

\subsection{Simulation Environment and Result}
The experimental evaluation conducts in OPNET Modeler 14.5, where several groups of SOs, two routers, the cloud, and two servers make up the testbed. In our testbed, WiFi clients, Wireless routers, and Ethernet servers simulate SOs, routers, and CA servers, respectively. We then reproduce registration with HTTP (heavy browsing) and authentication traffic with DB entry (medium load).

We first observe the average delay of registration and authentication if all traffic is sent and processed in the cloud (traditional cloud model). As outlined in Figure 4(a), the average registration delay increases from 0.19 second to 7.72 seconds when the node number varies from 10 to 120 pieces. Likewise, Figure 4(b) illustrates that the authentication traffic grows 40 times when the entity number jumps from 10 to 40 pieces. This sharp growth is further evidenced by Figure 4(c), where the authentication delay grows to more than 1000 seconds linearly. Unfortunately, the high authentication delay will trigger the failure of corresponding applications. To sum up, scalability and availability are poor when all data are sent and processed in the cloud.

\begin{figure*}
\centering
%\advance\leftskip-0.5cm
\includegraphics[width=0.9\textwidth]{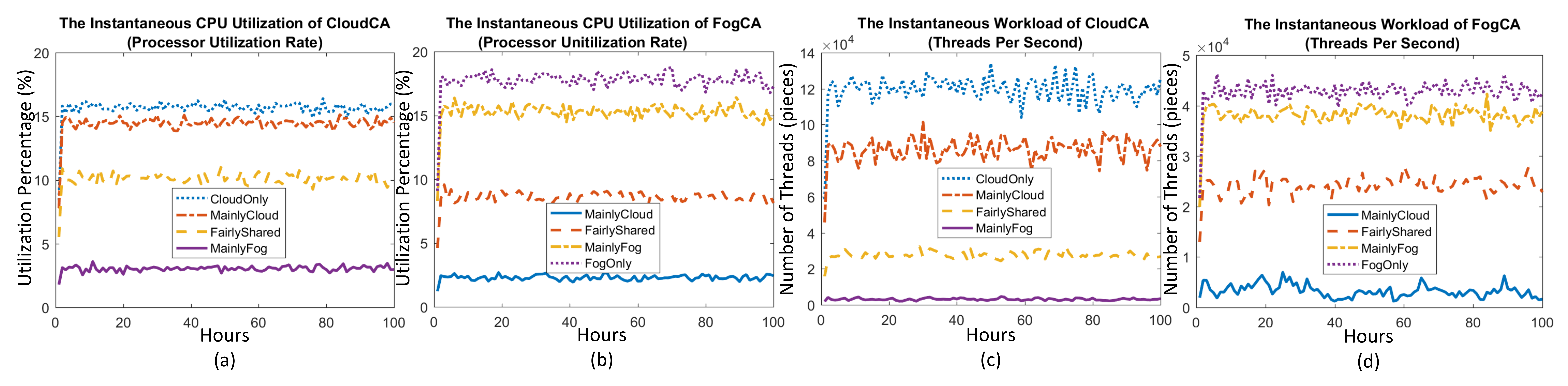}
\caption{The CPU utilization and workload in the shared model: (a) The instantaneous CPU utilization of CloudCA. (b) The instantaneous CPU utilization of FogCA. (c) The instantaneous workload of CloudCA. (d) The instantaneous workload of FogCA.}
\label{f6}
\end{figure*}

Different from the above cloud model that requires all data are sent and processed in a remote cloud, our CAaaS allows us to process the authentication traffic locally, thus decreasing the delay sharply. We define five settings to reflect such transition, i.e., CloudOnly (processed only in the cloud), MainlyCloud (90\% processed in the cloud, 10\% in the fog), FairlyShared (50\% in the cloud and 50\% in the fog), MainlyFog (10\% in the cloud, others in the fog), and FogOnly (processed only in the fog). As depicted in Figure 5(a) and 5(b), the registration delay is much reduced from above 5 seconds (CloudOnly) to under 0.2 seconds. More interestingly, even fog only processes 10\% of the registration traffic, the delay drops by about 42\%. Similarly, the CAaaS sharply diminishes authentication delay, as rendered in Figure 5(c) and 5(d). Indeed, the authentication delay in the FogOnly is about 1.7\% of that in the CloudOnly mode. 

Next, we review the CPU utilization and workload on both servers. To this end, Figure 6 (a) details the CPU utilization of the CloudCA while its fog counterpart is described in Figure 6(b). In general, CPU utilization grows with more pieces of work. Admittedly, the more conducted in the cloud, the fewer remains in the fog, and vice versa. The computing power is, actually, not a bottleneck in either model, as the utilization rate is less than 20\% in both servers.

Figure 6 (c) and (d) show the respective workloads on the two servers.  Again, the more processed in the cloud, the less is with fog. It is worth noting that the overall tasks on the FogCA are about 45 thousand per second, while on the CloudCA, it is about 140 thousand per second for the same CA services. The difference is due to the excessive data transmission incurred by network congestion. In more detail, when the end-to-end transmission delay is too high, the endpoint has to resend registration and authentication data to remote servers. Such retransmission led to the growth of traffic volume, also causing further delays. Again, the high latency of registration and authentication reduces the scalability and availability of CA services.

In summary, by distributing the required authentication and registration services to a \textit{Custodian} FN, the CAaaS much reduces latency and improves efficiency. Thereby, we showcase the benefits and success of computation movement to the proximity of data sources in the accomplishment of complicated applications.

\section{The Protocol Evaluation}
In this section, we conduct security analysis and formal verification to confirm that the CAaaS is effective against various attacks.

\subsection{Security Analysis}
The security analysis focuses on physical attack, passive attack, replay attack, and impersonation attack.

\subsubsection{Physical Attack}
Since fog cannot provide physical protection (e.g., robbery) for devices, in an insecure zone, we focus on detecting the unregistered modification when an SO joins fog. During the process, fog builds the situation and security awareness of SOs. As scrutinized in \cite{Li2018vfog}, fog manages and oversees an SO at its proximity. An SO informs a \textit{Custodian} of its original settings (ID, OS, \textit{IVV}, etc.). Then, the \textit{Custodian} verifies such information by taking advantage of the \textit{Parent-Child affinity}. The SO is trusted when the \textit{IVV} holds.

If a node has any modification after initial registration, the CAaaS can identify the node quickly according to a recalculated \textit{IVV} value. Under this circumstance, it will take further security action (such as device quarantine) to maintain the trustworthiness.

\subsubsection{Passive Attack}
Imagine that an attacker eavesdrops on the session between client and server, trying to derive the secrets in the system. With enough time and endeavors, the attacker may get $\{ID_C, M_C, R_C^*, T_1\}$, and $\{M_S, M_k, T_2\}$, and start to analyze the authentication key $A_{ID_C}$ and the server's private key $K_S^R$. At this instant, it needs to obtain the two values from either $M_C=R_C+t_1\times A_{ID_C}$ or $M_S=R_S+t_2\times K_S^R\times Q_{ID_C}$. The attacker would find it too difficult to achieve this because of the difficulties 1) and 3) presented in Section 3.1. For the same reason, the attacker is unable to figure out the session key from $M_k=(k+x_S)\times P$.

\subsubsection{Replay Attack}
Now we suppose that an attacker obtains the data being transferred between client and server, and pretends to be a legitimate client $C$ with $\{M_C, R_C^*\}$. Then, it sends $\{ID_C, M_C, R_C^*, T_1'\}$ to the server, where $T_1'$ is the current time. Since $M_C=R_C+t_1\times A_{ID_C}=R_C+H_2(T_1)\times A_{ID_C}$, while $T_1' \neq T_1$, the server will find that the verification equation in Step 2 of mutual authentication does not hold, that is, $R_C^*=x_C' \times P$. It is because of $M_C - K_S^R \times t_1' \times Q_{ID_C} \neq M_C-K_S^R\times t_1 \times Q_{ID_C}$, leading to $R_C' \neq R_C$. Vice versa, because of the past timestamp T2 on the client-side, the attacker is unable to claim to be the server by using $\{M_S, M_k\}$. 

\subsubsection{Impersonation Attack}
At this moment, we consider that an attacker claims to be a legal client C. It selects a point $R_C''=(x_C'',y_C'')\in E_p(a,b)$ and A$_{ID_C}''$ to calculate $M_C''=R_C''+t_1 \times A_{ID_C}''$ and ${R_C^*}''=x_C''\times P$. After this, $\{ID_C, M_C'', {R_C^*}'', T_1'\}$ is sent to the server for authentication. Since $M_C''$ is generated from $A_{ID_C}''$, rather than $A_{ID_C}$, the server is unable to get $R_C''=(x_C'',y_C'')$ from $R_C'=M_C''-K_S^R\times t_1\times Q_{ID_C}\neq M_C''-t_1\times A_{ID_C}$. Meanwhile, the server finds $R_C'' \neq x_C'\times P$, then determines that the attacker is an illegal user. Likewise, an attacker cannot present as a valid server without knowing the server's private key $K_S^R$.

\subsection{The Formal Verification Using AVISPA}

We use AVISPA to perform the verification, which is a popular evaluation tool for creating and troubleshooting formal models of security protocols \cite{CHALLA2020design}, \cite{WAZID2019design}. In short, we verify the confidentiality of authentication and key agreement message against intruders with a variety of attacking capabilities defined in the threat model. More significantly, AVISPA was developed based on the Dolev-Yao intruder model \cite{cervesato2001dolev}, where the intruder is allowed to intercept, inject, analyze, and change messages in transit at will. On top of this, we assess mutual authentication between clients (\textit{Child}) and servers (\textit{Custodian} and CA). Figure 7(a) further demonstrates our goal.

During the evaluation, we select the OFMC back-end to perform an automatic diagnosis and generate results. The result confirms that there is no major attack identified in the protocol. Our protocols are, therefore, secure against the active and passive attacks mentioned above. Figure 7(b) displays the detailed result.

To summarize, our proposed scheme is efficient (low-latency and scalable) and effective (against various attacks) for authenticating resource-constrained devices in the IoT.

\begin{figure}
\centering
\includegraphics[width=0.8\columnwidth]{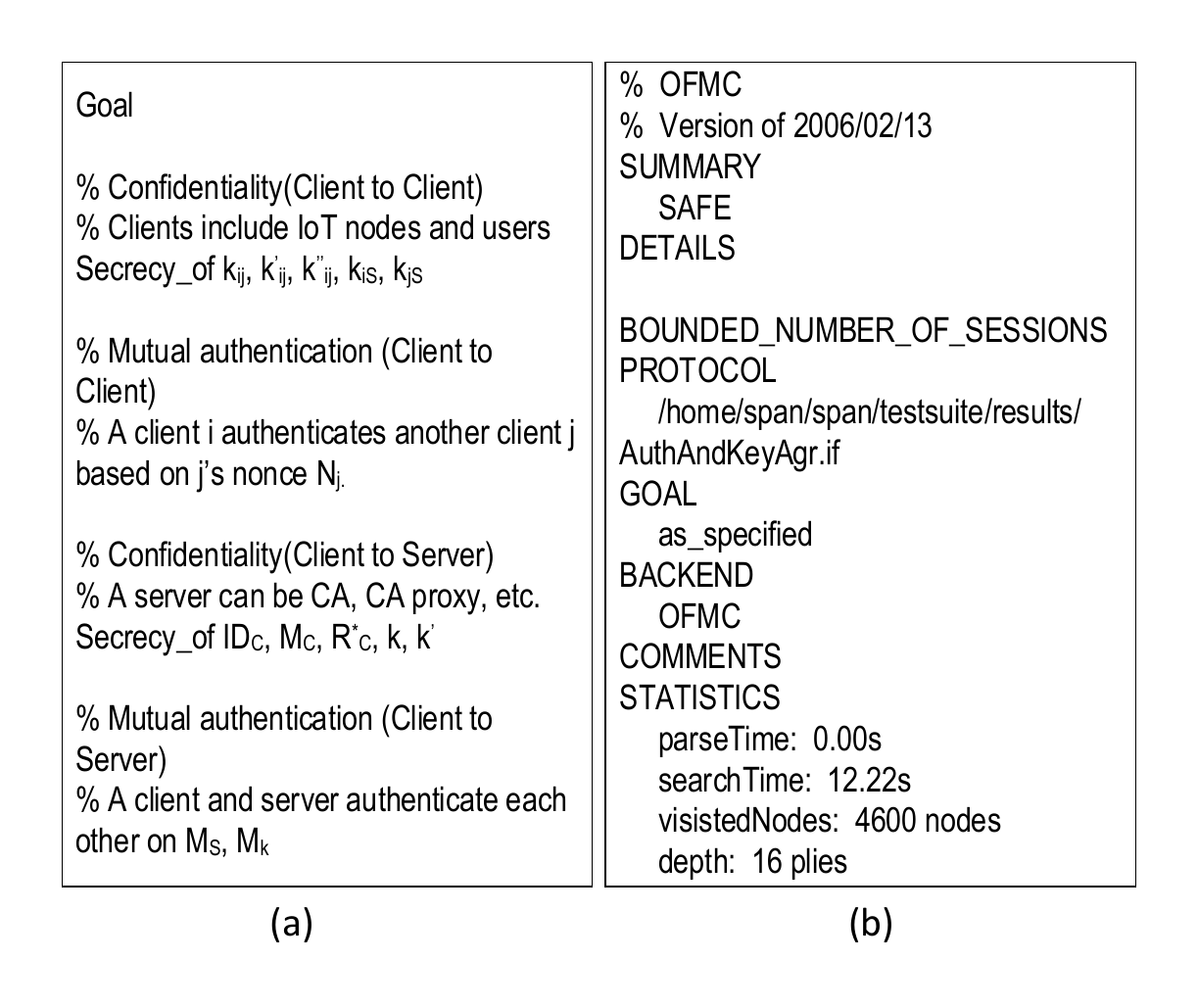}
\caption{The AVISPA evaluation. (a) The specific goal. (b) The result.}
\label{f7}
\end{figure}

\section{Related Work}
Authentication is a cornerstone in the construction of a security pillar \cite{jing2014security}, identifying the legitimacy of users and devices. To explain, Ali \textit{et al.} conducted a thorough study of security issues in respective layers of IoT architecture, claiming that a strong device authentication is critical in the perception layer \cite{ali2019internet}. Likely, Zhou \textit{et al.} stressed various attacks on pervasive user devices \cite{Zhou2015secure}. Reasonably, both academia and industry made impressive efforts in designing and implementing effective schemes for authenticating IoT devices. Silva \textit{et al.} surveyed main challenges for authentication in the IoT \cite{Silva2017authentication}, including low computation overhead, protocol deployment on resource-limited nodes, key management, and device heterogeneity. Likewise, Yang \textit{et al.} in \cite{Yang2017survey} reviewed relevant limitations, possible attacks, and solutions in IoT authentication, reiterating the necessity of a lightweight protocol for overcoming a diversity of resource constraints.

The research community recently proposed a myriad of cutting-edge concepts for IoT authentication, e.g., machine learning (ML), AI, big data \cite{AHMED2017bigdata}, 5G \cite{HUANG202010effective}, SDN, and blockchain. It is worth mentioning that neither ML, AI, big data, nor 5G is an independent technique in providing authentication, although these technologies have witnessed tremendous success in many areas. For example, Kaur \textit{et al.} applied big data in optimizing energy utilization for the IoT in \cite{kaur2020energy}. When these technologies are appropriately in use, they bring critical benefits in securing the IoT. To enumerate, Neshenko \textit{et al.} investigated various security issues and current solutions in their comprehensive survey paper \cite{Neshenko2019survey}, pinpointing the future initiative for the combination of technological advancements in addressing various security concerns raised in the IoT ecosystem. Garg \textit{et al.} developed a deep learning-based model for anomaly detection in \cite{garg2019depplearning}, improving accuracy and detection rate in securing cloud servers. Next, we briefly explore SDN and blockchain, respectively.

Shu \textit{et al.} investigated diverse threats and countermeasures for SDN in \cite{shu2016security}, arguing that SDN is capable of empowering security against various attacks. Nevertheless, the authors inadequately presented a related authentication scheme. Additionally, Salman \textit{et al.} introduced SDN in their identity-based authentication framework, claiming their approach takes control of various attacks such as replay attack and man-in-the-middle (MITM) attack \cite{Salman2016SDN}. Although this may be true, it was merely a conceptual framework without realization protocols. Furthermore, the authors did not study a things-to-things authentication approach in their work. In contrast, we provide a concrete solution for IoT authentications.

Many scholars proposed the blockchain technology in enabling the security for IoT as surveyed in \cite{KHAN2018blockchain}. Almadhoun \textit{et al.} further proposed a user authentication scheme for fog-based IoT in \cite{Almadhoun2018blockchain} where the blockchain technique was employed to protect fog nodes, via which, authenticating users to access IoT devices. Similarly, in \cite{Li2018blockchain}, Li \textit{et al.} came up with a blockchain-based mechanism to authenticate IoT devices, where a requested node searched the public key of the requesting node in the local blockchain or through consensus nodes recursively. Their scheme offered benefits such as resisting DDoS attacks and preventing malicious accesses. However, the performance would be heavily relying on the blockchain platform. In more detail, it is challenging to design and implement a contiguous blockchain platform in a highly dynamic environment and meanwhile reducing the search time.

Benefiting from a combination of passwords, smart cards, and biometrics, the multi-factor authentication (MFA) approach  \cite{DAS2018taxonmy} is becoming popular for mobile IoT devices \cite{ferrag2019authentication}. Compared with a single-factor system, it provides an additional tier of security in case one factor becomes compromised. Other benefits include ease of use, better compliance with related regulation, and improved user experience. Because such approaches usually need manual intervention, it is infeasible to numerous clientless things.

Situational awareness is the coordinated perception and comprehension of environmental elements in a network, which has played an important role in cybersecurity. In \cite{RAPUZZI2018building}, Rapuzzi \textit{et al.} proposed a three-layer architecture to establish the cognition for security threats, leveraging the technological combination of big data, SDN, NFV, fog, and cloud computing. Xiang \textit{et al.} developed a situation-aware authentication scheme for smart meters in \cite{xiang2020situation}, adjusting security protocol based on assessed security risk levels. When security risk is low, it conducts the low-cost protocol for optimized efficiency. Unfortunately, it is an open issue to assess security levels.

On the other hand, Ammar \textit{et al.} surveyed the security of mainstream IoT frameworks in the market provided by industrial giants, including Google, Amazon, Apple, and Microsoft. It seems that the industry still prefers CA-based solutions in the IoT authentication \cite{Mahmoud2018internet}. In summary, all industrial leaders scrutinized in the paper employed a CA-based PKI mechanism, which motivated us to focus on the CA-based authentication scheme in our study. Consequently, our solution may be more practical as it is in alignment with the mainstream authentication provisioning in today's market.

Note that a lightweight protocol is effective in addressing the earlier mentioned concerns. Garg \textit{et al.} put forward a lightweight authentication protocol for IoT in \cite{garg2020toward}, taking advantage of physically unclonable functions and XOR. In \cite{Salami2016Lightweight}, Salami \textit{et al.} designed a lightweight encryption scheme for the smart home, confirmed its efficiency in protecting smart environments. Due to its lightweight feature, ECC, to a large extent, has been proven its feasibility in IoT and sensor domain \cite{Li2018ECC}. He \textit{et al.} surveyed over thirty ECC-based authentication schemes satisfying the lightweight feature, while pointing out such schemes were insufficient in defending against replay attack, impersonation attack, and MITM attack in \cite{He2015ECC}. Lee \textit{et al.} designed a new authentication protocol, named ECDLP-based randomized access control \cite{Lee2008ECC}, in which the authors minimized the computational workload and the size of required memory. It was, however, incapable of handling impersonation attacks, as identified in \cite{bringer2008cryptanalysis}. In comparison, our protocol tackles such concerns.

It is as well beneficial to expedite security processing for enhanced security by employing a combination of software and hardware advancements. For example, Kotamsetty \textit{et al.}  proposed an adaptive latency-aware query processing approach over encrypted data in \cite{Kotamsetty2016adaptive}. In this work, the authors showed the latency drop by fine-tuning data size but did not evaluate security in their paper. In contrast, Liu \textit{et al.} proposed a hardware-accelerated ECC endomorphism, reducing delay up to 50\% of its original value as tested on the specially designed hardware platform in  \cite{Liu2017hardware}. However, products in the current market cannot take advantage of it.

Different from the above methods in authentication acceleration, our scheme takes advantage of virtual fog through which an authentication service is placed closer to users and devices, thus reducing the latency. To this end, computing intelligence moves closer to data instead of data going to remote servers in the cloud. Moreover, benefiting from \textit{Parent-Child affinity}, we minimize the attack surface on endpoints. As a result, our scheme is effective and efficient for IoT nodes, without much modification requirements of hardware and software. Meanwhile, we embrace the cloud-fog-things continuum in the CAaaS, which makes it more suitable for the IoT ecosystem. Last but not least, our solution protects previous investments in a progressive process in the construction of smart environments.

\section{Discussion}
Security is one of the most important characters that an established network must offer. Due to its inborn constraints of computing intelligence, many devices have insufficient power to safeguard themselves \cite{Neshenko2019survey}. It is fundamental to save incapable nodes with in-network protection intelligence, as a security level lies with an endpoint of the least security. To this end, our CAaaS steps in, just in time, by providing an effective and efficient authentication solution.

Powered by MSB, VNF, VO, and SDN, the VFOG is the foundation of our CAaaS scheme, through which, CA services step closer to data sources, resulting in low latency and high scalability. The VFOG framework brings local situational awareness along the cloud-fog-things continuum, contributing to the improved efficiency and effectiveness for authentication. It is worth noting that the VFOG is open to recent technological advancements. The collaboration of modern technologies will facilitate strengthening situational awareness for \textit{child}, \textit{parent}, and \textit{custodian}. However, it remains an open issue to concretely build situational security-awareness for resource-constrained IoT nodes with the current technique.

In this paper, we evaluate the authentication scheme using OPNET Modeler. In detail, we presented the end-to-end packet delay and workload on the server-side. In future evaluations, it is possible to contain more parameters such as the number of packets sent and received, and packet size information. Such data may aid in studying the simulation accuracy in comparison with data in an established work. As a reference for other scholars, it is also valuable to compare network diagrams and simulation methodology in future work.

The protocol to validate the CAaaS scheme is grounded in ECC. However, there are well-known security issues for the implementation of ECC, including side-channel attacks, possible NSA backdoors, and invalid curve attacks. Fan \textit{ et al.} concluded the leakage of the private key in several cases \cite{fan2010ecc}. It is, therefore, compulsory to perform a thorough security code review, static code evaluation, and stringent penetration testing to prevent improper implementation.

\section{Conclusion}
To conclude, we demonstrate the benefits of computing intelligence movements towards data sources in handling complicated services conducted traditionally in a cloud. In the proposed scheme, a variety of CA services encompassing key generation, storage, distribution, CRL, and CA proxies, separately performed on FN and reassembled to achieve required CA functions for authentication. Compared with a centralized cloud model requiring a large volume of data sent to and processed on a remote server, our scheme leverages fog computing and microservices, extending the intelligence of security computation to data sources with situational awareness. Evaluation results demonstrate the benefits of our CAaaS, including security, low latency, and high scalability. More interestingly, in the implementation of our protocols, computation-intensive workloads such as key generation are performed in fog, while resource-constrained devices only need to perform low-cost computation. We investigate the \textit{Parent-Child affinity} as a valuable input toward end-to-end security. Meanwhile, protocol evaluations confirm the effectiveness of anti-popular attacks. Our scheme is more suitable to resource-limited entities in smart living and other similar environments, grounded in benefits such as efficiency and effectiveness without computation-intensive workload requirements on the client-side.

\bibliographystyle{elsarticle-num}
%\bibliographystyle{cas-model2-names}

\bibliography{main}


\bio{}
Jianhua Li received the  Ph.D. degree in Information Engineering from Swinburne University of Technology, Melbourne, VIC, Australia, in 2019. He is currently a Lecturer with the School of Information Technology, Deakin University, Melbourne, VIC, Australia. Meanwhile, he is an ICT professional with extensive hands-on experience on multi-million dollar projects. His current research interests include fog computing, the Internet of Things, quality of experience, artificial intelligence, cybersecurity, and networking in operational projects.
\endbio

%\bio{figs/pic1}
\bio{}
Jiong Jin (M'11) received the B.E. degree with First Class Honours in Computer Engineering from Nanyang Technological University, Singapore, in 2006, and the Ph.D. degree in Electrical and Electronic Engineering from the University of Melbourne, Australia, in 2011. From 2011 to 2013, he was a Research Fellow in the Department of Electrical and Electronic Engineering at the University of Melbourne. He is currently a Senior Lecturer in the School of Software and Electrical Engineering, Faculty of Science, Engineering and Technology, Swinburne University of Technology, Melbourne, Australia. His research interests include network design and optimization, edge computing and distributed systems, robotics and automation, and cyber-physical systems and Internet of Things.
\endbio

\bio{}
Lingjuan Lyu (M'18) is currently a Research Fellow with The Department of Computer Science, National University of Singapore. She received Ph.D. degree from the University of Melbourne. Her current research interests span machine learning, privacy, fairness, and edge intelligence. She has publications in prestigious venues, including: JSAC, TPDS, TDSC, TCC, JIOT, TII, DKE, SIGIR, WWW, CIKM,  FL-IJCAI, IJCNN, TrustCom, PerCom, etc. Her work was supported by an IBM Ph.D. Fellowship (50 winners Worldwide) and contributed to various professional activities, including AISTATS, ICML, NIPS, NIPS workshop, S\&P, PETS, JIOT, TCC, TII, Neurocomputing, ACM Computing Surveys, TKDE, TDSC, TNSM, IEEE Access etc.
\endbio

\bio{}
Dong Yuan received the B.E. and M.E. degrees from Shandong University, Jinan, China, in 2005 and 2008, respectively, and the Ph.D. degree from the Swinburne University of Technology, Melbourne, VIC, Australia, in 2012. He is a Lecturer with the School of Electrical and Information Engineering, University of Sydney, Sydney, NSW, Australia. His current research interests include cloud computing, data management in parallel and distributed systems, scheduling and resource management, Internet of Things, business process management, and workflow systems.
\endbio

\bio{}
Yingying Yang  received the B.E in Harbin Institute of Technology, Harbin, China in 2002 and M.E degrees from Charles Sturt University, Sydney, NSW, Australia in 2013. She is a current PhD student in University Technology Sydney, NSW, Australia. She is a distinguished CISCO Networking Academy Instructor trainer with more than twelve years’ experience. Her current research interests include cloud computing, fog/edge computing, physical-cyber systems, Internet of Things, and software-defined networking.
\endbio

\bio{}
Longxiang Gao received his PhD in Computer Science from Deakin University, Australia. He is currently a Senior Lecturer at School of Information Technology, Deakin University. His research interests include fog/edge computing, blockchain, data analysis and privacy protection. Dr. Gao has over 80 publications, including patent, monograph, book chapter, journal and conference papers. Some of his publications have been published in the top venue, such as IEEE TMC, IEEE IoTJ, IEEE TDSC, IEEE TVT, IEEE TCSS, and IEEE TNSE. He is a Senior Member of IEEE.
\endbio

\bio{}
[S'09, M'14, SM'20] is a full Professor in the faculty of Electronic and Information Engineering, Xi'an Jiaotong University of China. He was a research scholar in Computer Science Department at Carnegie Mellon University from 2011 to 2013. His current research interests mainly include Data-Driven Network and System Security, AI Security, Cyber-Physical System Security. He currently serves as an Associate Editor for a number of journals, including IEEE Transactions on Dependable Secure Computing, Journal of Franklin Institute, and Frontiers of Computer Science, etc. 
\endbio

\end{document}